# A Controllable and Highly Propagative Hybrid Surface Plasmon-Phonon Polariton in a CdZnO-based Two-Interface System

J. Tamayo-Arriola[1,*], E. Martínez Castellano[1,*], M. Montes Bajo[1], A. Huerta-Barberà[2], E. Muñoz[1], V. Muñoz-Sanjosé[2], A. Hierro[1,†]

[1] ISOM, Universidad Politécnica de Madrid, Madrid, Spain

[2] Dept. Física Aplicada i Electromagnetisme, Universitat de València, Burjassot, Spain

## Abstract

The development of new nanophotonic devices requires the understanding and modulation of the propagating surface plasmon and phonon modes arising in plasmonic and polar dielectric materials, respectively. Here we explore the CdZnO alloy as a plasmonic material, with a tunable plasma frequency and reduced losses compared to pure CdO. By means of attenuated total reflectance, we experimentally observe the hybridization of the surface plasmon polariton (SPP) with the surface phonon polariton (SPhP) in the air-CdZnO-sapphire three-layer system. We show how through the precise control of the CdZnO thickness, the resonance frequencies of the hybrid surface plasmon-phonon polariton (SPPP) are tuned in the mid-infrared, and the nature of the hybrid mode turns from a plasmon-like behavior in the thicker films to a phonon-like behavior in the thinnest films. The presence of sapphire phonons not only allows the hybrid mode to be formed, but also improves its characteristics with respect to the bare SPP. The reduced damping of the phonon oscillators allows to reduce the losses of the hybrid mode, enhancing the propagation length above 500 μm, one order of magnitude larger than that of typical SPPs, clearing the path for its application on emerging devices such as plasmonic waveguides.

---

[*] These authors contributed equally to this work

[†] Corresponding author: adrian.hierro@upm.es

The outstanding properties of electromagnetic surface waves and the recent progress in the synthesis of semiconductors with optimized properties, has renewed the interest in the field of plasmonics and its application to new nanophotonic devices.[1–3] As a result, there is a growing demand of functional photonic devices based on light-matter interactions at interfaces that can overcome the limitations of the current technology of semiconductor integrated circuits,[4] exploring the light confinement phenomenon beyond the diffraction limit.[5]

Light interaction with doped semiconductors and polar dielectrics leads to the formation of surface plasmon polaritons (SPPs) and surface phonon polaritons (SPhPs), respectively. In such media, when the thickness is reduced to be comparable to the optical skin depth, epsilon-near-zero (ENZ) modes arise, at the frequency where the dielectric function vanishes.[6] While ENZ modes have the extraordinary ability to confine light within sub-wavelength thin films,[7] and therefore are suitable in photovoltaics and bolometer devices[2] and for nonlinear optics,[3] they lack the property of being dispersive, and thus show very low group velocities and propagation lengths. Conversely, SPPs and SPhPs have lower light confinements, but much larger group velocities, a key point when signal transport along an interface is required, as in plasmonic waveguides.[4] In order to further exploit this advantage, here we experimentally and theoretically explore the hybridization of SPPs with SPhPs in the air-CdZnO-sapphire three-layer system with very low optical losses, looking for the enhancement of the propagation length of the hybrid mode as compared to the bare SPP.

The excitation of SPPs in the mid-infrared (IR) requires materials with carrier concentrations between $10^{19}$ to $10^{21}$ cm$^{-3}$, as well as high electron mobilities and low optical losses. Thus, metals are discarded due to their very high electron concentrations and high optical losses, and transparent conductive oxides (TCOs) become the best alternative.[8] However, typical TCOs such as ITO and ZnO are far from being ideal for mid-IR plasmonics due to their high plasma dampings ($\sim 800 - 900 \ cm^{-1}$),[9,10] which result in low electron mobilities. In this sense, CdO fulfills the requirements and is postulated to be the best candidate among all TCOs, taking advantage of its high electron concentrations and high electron mobilities in as-grown material.[11,12] In addition, the electron concentration, and therefore the plasma frequency can be modulated by doping or alloying it with other materials.[13–16] For instance, it has recently been shown that through the controlled doping of CdO with F the position of the ENZ resonance can be tuned in a range of frequencies from 1800 to 3668 cm$^{-1}$.[17] Alternatively, we have proposed the rock-salt CdZnO alloy for tuning the plasma frequency in the mid-IR, between 3000 and 4000 cm$^{-1}$. Alloying CdO with ZnO in the rock-salt phase also allows us to improve the Hall electron mobility, yielding values as high as 110 cm$^2$/V·s.[18]

In this study we present a detailed analysis of the hybridization of two distinct-in-nature surface modes: the SPP, produced by CdZnO, a plasmonic oxide, and the SPhP, produced by sapphire, a polar dielectric crystal, which also serves as the substrate where the CdZnO film is epitaxially grown. The resulting hybrid mode, the surface plasmon-phonon polariton (SPPP), first proposed by M. Nakayama,[19] takes the advantages of the bare phonon and plasmon surface modes, with a much higher propagation length and a wider range of attainable frequencies. The SPPP hybrid mode has been experimentally observed, among other 2D systems, at the graphene-boron nitride interface, confined in extremely small volumes.[20] Here we present the first experimental observation of SPPPs in oxides, much easier to synthesize and with extraordinary electrical properties for their implementation in the aforementioned plasmonic devices. Moreover, as compared to a graphene plasmonic monolayer, the thickness and plasma frequency of thin film oxides can be easily controlled, allowing to tune the optical response of the SPPP modes over a wide range of frequencies in the mid-IR.

In the air-CdZnO-sapphire three-layer system here explored, the coupling strength of the fields at the air-CdZnO and CdZnO-sapphire interfaces is easily controlled by the CdZnO thickness. This coupling leads to the well-known symmetric and antisymmetric field distributions, which appear at different energies.[7,21,22] Note that our system is not perfectly symmetric, since the dielectric functions of air and sapphire are different, but we will refer to the symmetry of the sign of the field distributions, for simplicity. The most energetic mode, i.e. the symmetric mode, shifts to higher energies when the CdZnO thickness is reduced, and it is pinned at the plasma frequency in the extremely thin case. It corresponds to the ENZ mode, where the dielectric function of CdZnO vanishes. The less energetic mode, i.e. the antisymmetric mode, which in fact is the SPPP, turns from a plasmon-like behavior for frequencies far from the sapphire phonons in the thickest CdZnO film to a phonon-like behavior as the thickness of the film is reduced and the coupling of the fields at both interfaces becomes stronger. Precisely, in the thinner CdZnO films the effect of phonons is more evident, with a direct effect on the enhancement of the propagation length of the SPPP mode. In the ultrathin limit case, the surface mode is dominated by the sapphire SPhP. Sapphire has several phonon modes but the one hybridized with the oxide surface plasmon is the most energetic one, with its transversal optical (TO) mode at 633 cm$^{-1}$ and its longitudinal optical (LO) mode at 905 cm$^{-1}$ (see Supporting Information).

The $Cd_{1-x}Zn_xO$ films are grown by metal-organic chemical vapor deposition (MOCVD) on *r*-plane sapphire (see Ref. 18 for details), and have rock-salt crystalline structure and variable thicknesses ranging from 25 to 460 nm. The Zn content in the ternary alloy is set to be x=10 %, showing a plasma frequency around 4000 cm$^{-1}$ and larger electron mobilities than for pure CdO.[18] It is worth noting that the polaritons can be observed thanks to the high crystal quality achieved both in the CdZnO films as well as at the CdZnO-sapphire interface. The deposition of the film by MOCVD allows achieving these requirements and having a precise control of the CdZnO thickness, crucial to select the frequencies where the SPPP can be excited.

In order to excite the surface resonances, attenuated total reflectance (ATR) measurements were carried out in a Fourier Transform Infrared (FTIR) spectrometer, where the incident light was *p*-polarized and the in-plane momentum was matched to the polariton momentum, $k_x$, by using a ZnSe prism in the Otto configuration.[23] The air gap between the ZnSe prism and the CdZnO film was determined to be about 350 nm in all the experiments, small enough to excite the SPPP and large enough to nearly decouple the prism-air interface from the system. The angle of incidence ($\theta$) in the CdZnO film was controlled from 39° to 59°, allowing to scan the polariton dispersion curve through $k_x = k_0 \cdot n_p \cdot sin(\theta)$, where $n_p = 2.37$ is the prism refractive index and $k_0$ the light momentum in vacuum.

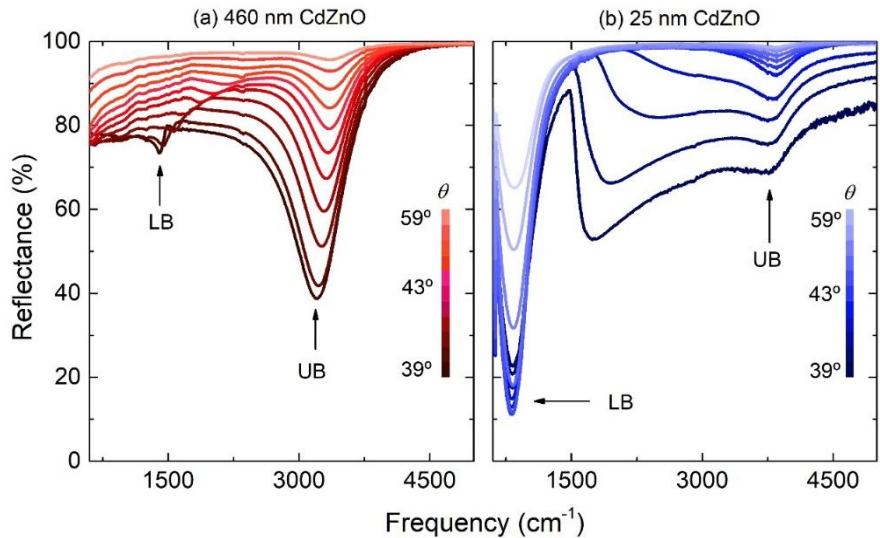

**Figure 1.** Measured attenuated total reflectance curves in *p*-polarization at different angles of incidence for the two extreme CdZnO thicknesses: (a) 460 nm and (b) 25 nm. The upper branch (UB) and the lower branch (LB) are indicated in each case.

The measured ATR curves of the two CdZnO extreme thicknesses, i.e. 460 and 25 nm, are shown in Figure 1. The measured and simulated ATR curves of all samples can be found in the Supporting Information. In the thickest CdZnO film (Figure 1 (a)), the air-CdZnO and the CdZnO-sapphire interfaces are almost fully decoupled, and the system behaves as having two independent interface layers, each with its surface mode: the SPP formed at the air-CdZnO interface corresponding to the upper branch (UB), and the SPPP formed at the CdZnO-sapphire interface corresponding to the lower branch (LB). In Figure 1 (a) it can be seen how the LB is almost negligible, owing to the exponential decay of the evanescent wave within the CdZnO, which hardly reaches the CdZnO-sapphire interface.

On the other hand, in the 25 nm-thick CdZnO film (Figure 1 (b)) the fields at both interfaces are coupled and arrange in a symmetric distribution for the SPP (UB), and in an asymmetric distribution for the hybrid SPPP (LB). As a result, the difference in energy of the two branches increases, and the reflectance drop in the asymmetric branch becomes narrower and more pronounced, while the symmetric branch losses prominence. This is a consequence of two facts: first, the thickness of the plasmonic material is reduced and so does the strength of its associated SPP; second, the phonon-like character acquired by the SPPP hybrid mode, which has a lower plasmon-phonon damping. Indeed, from the reflectance measurements (see Supporting Information) the damping of the plasma frequency was deduced to be 500 cm$^{-1}$ and the dampings of the TO and LO phonons 6 and 17 cm$^{-1}$, repectively Therefore, the damping of the hybrid mode is expected to vary between these two limits, depending on the proximity of the resonance to the sapphire phonons.

The overall evolution of the two branches with the thickness of the CdZnO layer can be observed in Figure 2, where the ATR contour plots of all the samples are shown. The plots were modelled by applying the Transfer Matrix Method (TMM) to the air-CdZnO-sapphire system with a ZnSe prism on top, and the symbols indicate the resonance frequencies experimentally obtained. The parameters used for modelling the ATR contour plots were previously deduced by means of conventional infrared reflectance measurements and can be found in the Supporting Information.

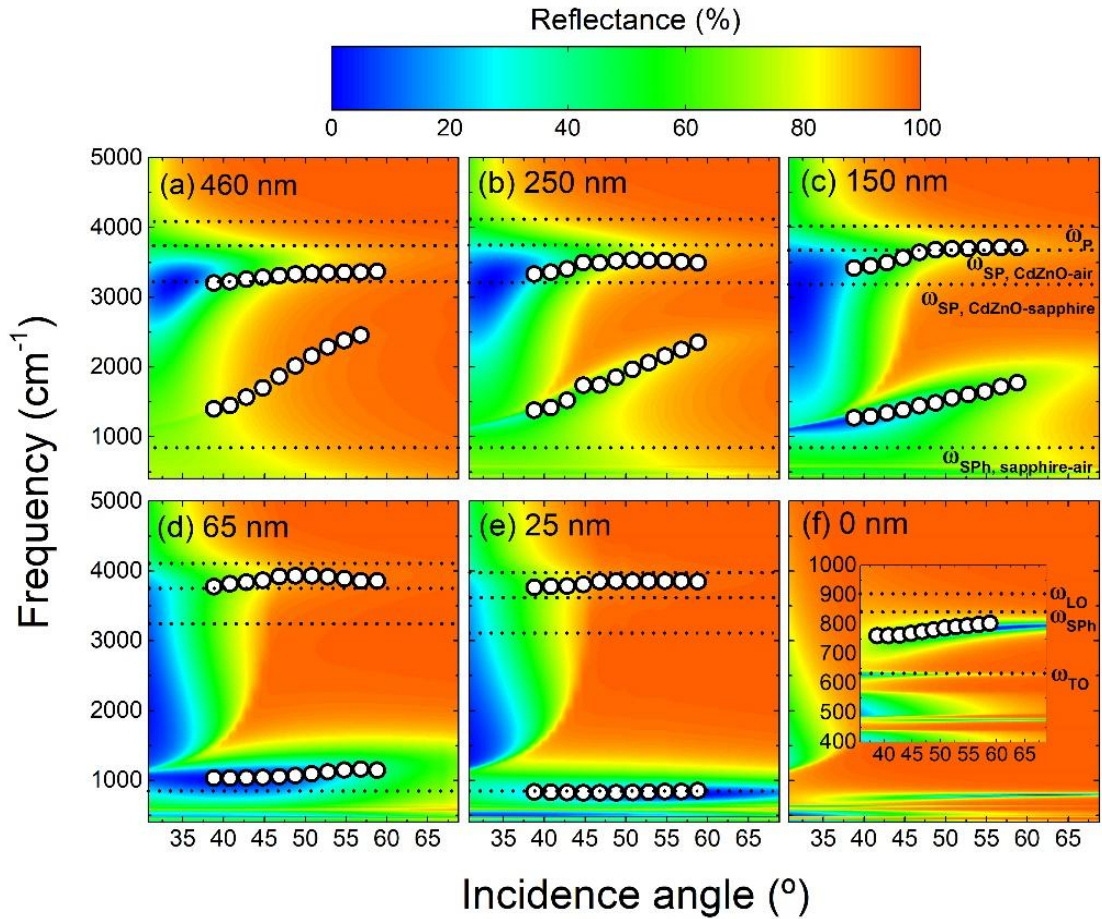

**Figure 2.** From (a) to (f), simulated ATR contour plots and experimentally determined resonance frequencies (circular dots) in *p*-polarization for the air-CdZnO-sapphire three-layer system, for each CdZnO thickness. The horizontally dotted lines indicate, from the highest to the lowest frequency: the plasma frequency ($\omega_p$), the asymptote of the bare surface plasmon formed at the CdZnO-air interface

($\omega_{sp,\ CdZnO-air}$), the asymptote of the bare surface plasmon formed at the CdZnO-sapphire interface ($\omega_{sp,\ CdZnO-sapphire}$), and the asymptote of the bare surface phonon formed at the sapphire-air interface ($\omega_{sph,\ sapphire-air}$). The inset in (f) shows in more detail the reststrahlen band of the most energetic sapphire phonon, where the SPhP is found. A comparison of the experimental and simulated ATR contour plots can be found in the Supporting Information.

Starting with the thickest CdZnO film (Figure 2 (a)), the UB is a pure plasmonic mode, and has an asymptote at $\omega_{sp,\ CdZnO-air} = \omega_p \cdot (1 + \varepsilon_\infty^{CdZnO})^{-1/2}$ when the in-plane momentum goes to infinity. On the other hand, although the LB is almost negligible, it corresponds to the hybrid SPPP mode, with a phonon-like behavior for lower in-plane momenta and a plasmon-like behavior with an asymptote at $\omega_{sp,\ CdZnO-sapphire} = \omega_p \cdot \left(\varepsilon_\infty^{Sapphire} + \varepsilon_\infty^{CdZnO}\right)^{-1/2}$ for higher in-plane momenta.

As can be observed in Figure 2 (b) to (e), as the thickness decreases the fields at both interfaces start to interact, resulting on the separation of the symmetric and antisymmetric branches. The symmetric one approaches the ENZ mode, where the dielectric function of the material vanishes at the plasma frequency, appearing only for thicknesses below the skin depth of the material. In fact, it can be seen how in our 25 nm CdZnO sample, the resonances of the UB are about 3850 cm$^{-1}$, very close to the plasma frequency, at 3970 cm$^{-1}$. If the thickness were further reduced, the resonances would be pinned at the plasma frequency. In contrast, the antisymmetric branch approaches the reststrahlen band formed within the sapphire TO and LO phonons for reduced thicknesses. Indeed, for the thinnest CdZnO film the behavior of the LB approaches that of the pure sapphire surface phonon polariton shown in Figure 2 (f), confined within the reststrahlen band, from the TO phonon for low in-plane momentum to the asymptote at $\omega_{sph} = 845\ cm^{-1}$ when the in-plane momentum tends to infinity. At the resonance, the fraction of the reflected light is close to zero, acting as a nearly-perfect absorber.

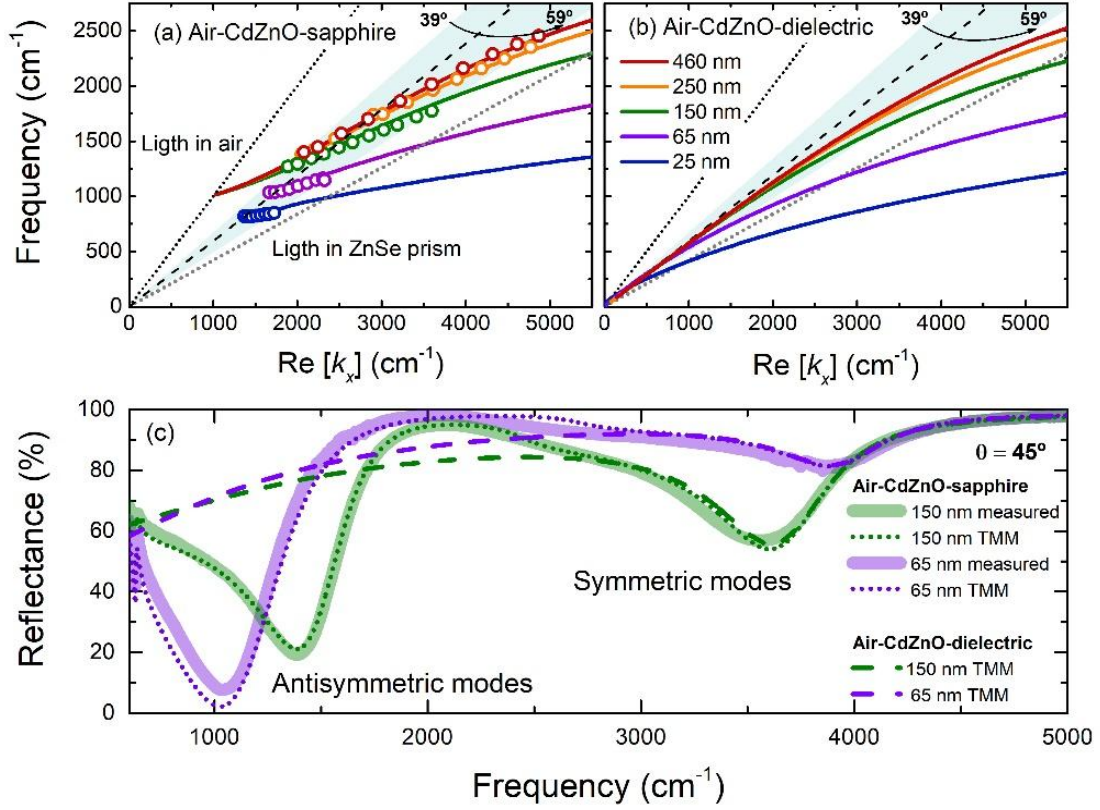

**Figure 3**. Lower branch dispersion curves computed with the TMM for all CdZnO thicknesses for (a) the air-CdZnO-sapphire system, where the circular symbols indicate the measured resonance positions, and (b) the idealized air-CdZnO-dielectric system, where the dielectric function of sapphire is free of phonon resonances. The shaded blue regions indicate the measurable range with the ZnSe prism, from an angle of incidence of 39º to 59º. The dashed line represents the light line for an angle of incidence of 45º with the ZnSe prism. (c) ATR measurements and simulations of the air-CdZnO-sapphire and idealized air-CdZnO-dielectric system, at an angle of incidence of 45º.

In order to properly evaluate the effect of sapphire phonons on the hybrid mode, the dispersion curves of our air-CdZnO-sapphire system with variable CdZnO thicknesses (Figure 3 (a)) are compared to that of an idealized system on which phonons have been removed from the sapphire dielectric function (Figure 3 (b)), i.e. the sapphire substrate is substituted by a dielectric with a constant dielectric function equal to 3, which indeed is the high-frequency dielectric constant of sapphire, $\varepsilon_\infty^{Sapphire}$. The range of frequencies where the LB is visible is clearly modified by the effect of sapphire phonons, as observed when comparing Figure 3 (a) and (b). Moreover, the LB is much more dispersive with a phonon-free dielectric substrate, yielding a pure SPP, and for small variations of the angle of incidence the shift of the resonance is considerable. This could indicate a higher propagation length ($L_p$) of the pure SPP mode along the *x*-axis compared to the hybrid SPPP, since the group velocity, defined as $v_g = \partial\omega/\partial Re[k_x]$, is higher. However, as discussed below, thanks to the reduced damping of the hybrid SPPP mode compared to the bare SPP, its losses are reduced and therefore its $L_p$ is largely enhanced.

The effect of the reduced damping is appreciable in the measured and simulated ATR curves shown in Figure 3 (c). In the air-CdZnO-sapphire system the antisymmetric mode is clearly seen, with a fraction of reflected light close to 0 %, especially when the resonance approaches the LO sapphire phonon. In contrast, for a phonon-free dielectric substrate the resonance is at much lower frequencies (Figure 3 (b)) and thus is not detectable in the mid-IR, as simulated with the dashed lines of Figure 3 (c). In order to make it visible in the mid-IR the thickest samples have to be chosen and the angle of incidence has to be higher than 52°. However, it is worth remembering that when the CdZnO thickness is increased, the strength of the LB is reduced since the evanescent wave hardly reaches the CdZnO-sapphire interface. Finally, regarding the symmetric mode, as observed in Figure 3 (c) the resonance is unaffected by phonons, since their respective frequencies are found far from each other.

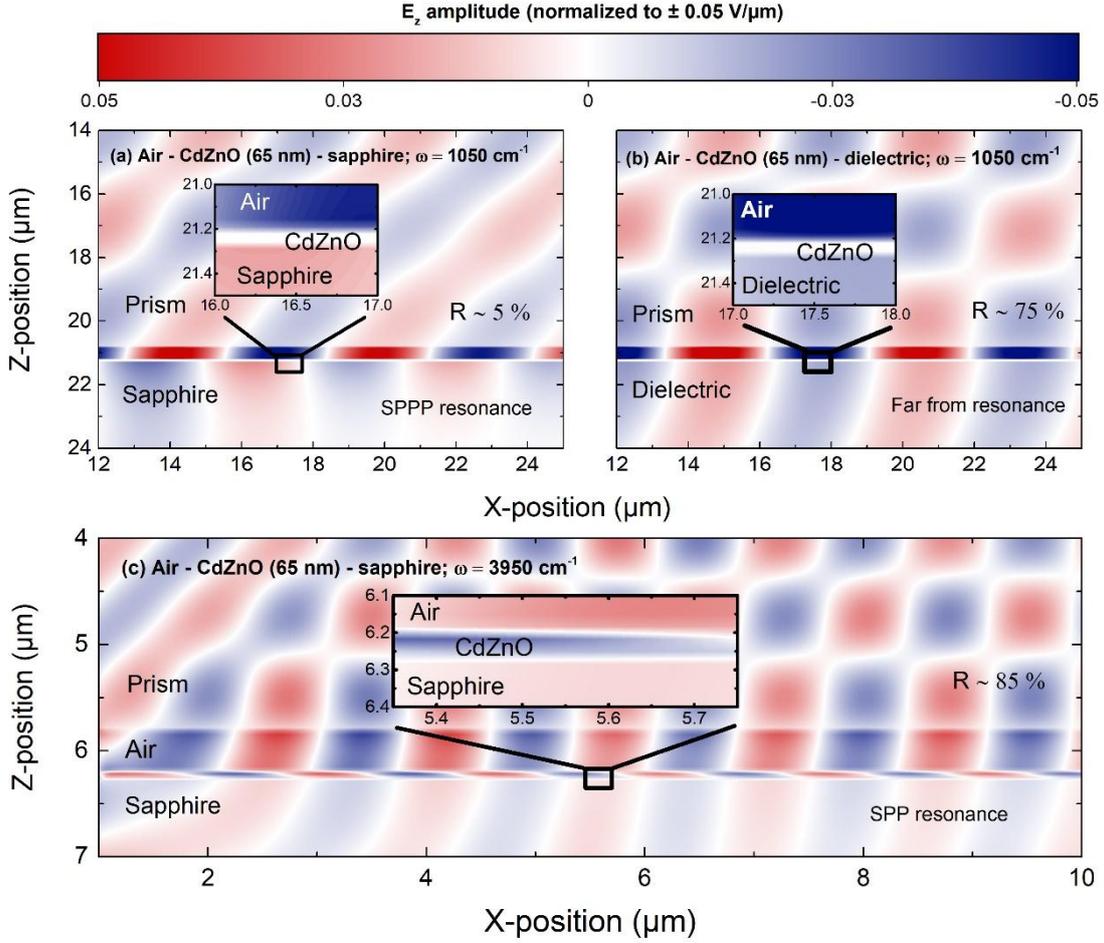

**Figure 4.** Distribution of the *z*-component of the electric field of a plane wave propagating along the (1,0,1) direction with a frequency of 1050 cm$^{-1}$, impinging at 45º on the air-CdZnO-sapphire system (a), and on the idealized air-CdZnO-dielectric system (b). In (c), the frequency of the plane wave is 3950 cm$^{-1}$ and it impinges in the air-CdZnO-sapphire system. For all cases, the inset shows a zoom of the region of interest.

To confirm the existence of the symmetric and antisymmetric modes, the *z*-component of the electric field ($E_z$) at the XZ-plane in the CdZnO film with a thickness of 65 nm was calculated, using the finite-difference time-domain (FDTD) method, from an open-source software package,[24] and is shown in Figure 4. Again, in order to evaluate the effect of phonons on the LB, the $E_z$ distribution in our system (Figure 4 (a)) is compared with that of an equivalent idealized system where the sapphire is substituted by the dielectric without phonons (Figure 4 (b)). As observed in Figure 4 (a), the field distribution is antisymmetric at the frequency where the SPPP is found, i.e. at 1050 cm$^{-1}$ for an angle of incidence of 45°, with opposite sign in the air and sapphire half-spaces. Within the CdZnO, $E_z$ is close to zero, as expected for an antisymmetric mode.[22] When the phonons in the substrate are removed, as in Figure 4 (b), the antisymmetric mode is dissipated, with the plane wave being almost unaltered by the dielectric. Besides, as expected from Figure 3 (b), the fraction of reflected light has now increased from 5 to 75 % since no light is absorbed by any resonance, and an interference pattern is formed in the prism.

When the frequency of the incident light matches the resonance of the UB, the $E_z$ distribution in Figure 4 (c) is formed. Now the field distribution is symmetric, with the same sign for $E_z$ in air and in the sapphire substrate, and opposite sign within the CdZnO thin film. When the thickness of the CdZnO film is further reduced to 25 nm and below, this resonance approaches the ENZ mode and the $E_z$ magnitude within the thin film is further enhanced.

Once the nature of the modes has been analyzed, it is worth evaluating whether the hybridation of the oxide SPP with the sapphire SPhP improves the key field parameters. To answer this question, and focusing on the SPPP hybrid mode, we have computed the propagation length ($L_p$) along the interface and the transverse field confinement distance ($\delta_z$) in air. The values obtained for the SPPPs of our set of samples are compared with those of the SPPs formed in the idealized phonon-free air-CdZnO-dielectric system and shown in Figure 5.

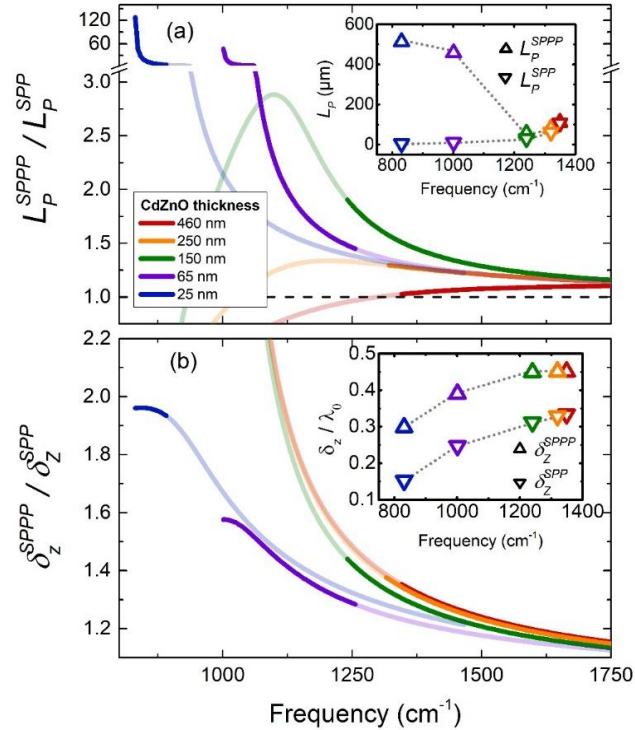

**Figure 5.** (a) Ratio of the propagation length along the *x*-direction between the SPPP and the phonon-free SPP vs. the frequency of the incoming light. Note the change in the *y*-scale at *y*=3.1. The inset shows the absolute values of $L_P^{SPPP}$ and $L_P^{SPP}$ for each sample, at the minimum measurable frequency in each case. (b) Ratio of the field confinement distance in air between the SPPP and the phonon-free SPP vs. the frequency

of the incoming light. The inset shows the confinement length as compared to the wavelength of the light, at the minimum measurable frequency in each case. The highlighted regions of the curves in both graphs correspond to the attainable frequencies with the ZnSe prism from an angle of incidence of 39º to 59º.

The propagation length is limited by the losses in the CdZnO and sapphire media, and therefore is inversely proportional to the imaginary part of the polariton momentum, as $L_p = 1/2Im\{k_x\}$. In Figure 5 (a), the ratio of the propagation length with and without phonons in the substrate as a function of the frequency of the incoming light is represented for every CdZnO thickness, i.e., the propagation length of the SPPP ($L_P^{SPPP}$) in the air-CdZnO-sapphire system is compared to that of the bare SPP ($L_P^{SPP}$) in the phonon-free air-CdZnO-dielectric system. In order to provide a reference, the range of frequencies where the polariton can be excited in our geometry is highlighted.

For a CdZnO thickness of 150 nm, the $L_P^{SPPP}/L_P^{SPP}$ ratio reaches a maximum of 2.9 at a frequency of 1100 cm$^{-1}$ and then it drops for lower frequencies. This is explained as follows: considering the dispersion curve in Figure 3 (b) corresponding to the thickness of 150 nm, below a frequency of ~1100 cm$^{-1}$ the SPP of the air-CdZnO-dielectric system has a nearly-constant slope, being in essence a lossless photon propagating along the interface. In contrast, as seen in Figure 3 (a) for the same frequencies and thickness, the SPPP of the air-CdZnO-sapphire system is not photon-like but phonon-plasmon-like with higher losses, explaining the drop of $L_P^{SPPP}$ as compared to $L_P^{SPP}$.

Conversely, for lower CdZnO thicknesses the frequencies at which the dispersion curves in Figure 3 (b) approach the photon-line are much lower. Besides, the resonances in Figure 3 (a) are closer to the reststrahlen band of the sapphire phonons, and the hybrid mode acquires a phonon-like character with reduced losses. This explains the sudden rise of $L_P^{SPPP}$ with a sapphire substrate when the frequency approaches the phonon frequencies. For a CdZnO thickness of 65 nm, $L_P^{SPPP}$ is 46 times $L_P^{SPP}$ at an angle of incidence of 39°, and for a thickness of 25 nm the ratio is 128. Considering the absolute values shown in the inset of Figure 5, the propagation length reaches the outstanding values of 513 μm for the 25 nm sample and 458 μm for the 65 nm sample, much higher than the typical values for SPPs in the mid-IR. Indeed, looking at the ENZ-SPP hybrid mode observed by Runnerstrom *et al.* in a CdO system,[25] the propagation lengths they obtained are in the range of 5-40 μm in the ultra-strong coupling regime. The propagation length is one order of magnitude larger in our study, and arises from the fact that the antisymmetric mode formed in our system does not penetrate the plasmonic material (see Figure 4 (a)) and so the SPPP mode suffers from lower losses than the ENZ-SPP mode. In a purely phononic structure, such as that reported by N.C. Passler for the air-AlN-SiC system,[26] and thanks to the reduced damping of phonons, the ENZ-SPhP hybrid mode has revealed to support a propagation length of 900 μm. However, it must be noticed that when the surface modes are entirely phononic in nature other shortcomings arise, such as the limited tunability of the resonance frequencies.

Regarding the transverse field confinement distance, which will dictate the degree of miniaturization of the components in a potential nanophotonic circuit,[27] it is inversely proportional to the imaginary part of the transverse component of the polariton momentum, as $\delta_z = 1/2Im\{k_z\}$. The confinement distance of the antisymmetric SPPP mode ($\delta_z^{SPPP}$) cannot be as low as that in of ENZ-modes, where most of the field intensity lies within the plasmonic material.[7] As observed in Figure 5 (b), the ratio of the field confinement distances with and without phonons in the substrate, i.e. the ratio between $\delta_z^{SPPP}$ and the confinement distance of the bare SPP ($\delta_z^{SPP}$), rises when the frequency is lowered, by the effect of sapphire phonons. However, its maximum value does not exceed 2 and 1.6 for the 25 nm and 65 nm CdZnO films, respectively, which again are the thicknesses for which the hybrid SPPP mode is better formed. We thus confirm the loss in confinement is negligible as compared to the gain in the propagation length discussed above. Indeed, looking at the values in the inset of Figure 5 (b), the confinement distance covers the range of $(0.15 - 0.30)\lambda_0$, where $\lambda_0$ is the wavelength of the incident light, for the system without phonons, and the range of $(0.30 - 0.45)\lambda_0$ for the system with phonons. Thus, these confinements are similar to those of typical SPPs.

In summary, in this work we have experimentally demonstrated the existence of a surface plasmon-phonon polariton hybrid mode in the air-CdZnO-sapphire three-layer system, which is observable thanks to the high crystal quality of the CdZnO films and their interface to sapphire, as well as their reduced optical losses. The coupling between the fields at the air-CdZnO and CdZnO-sapphire interfaces is shown to be controllable through the CdZnO thickness, and they arrange into the symmetric ENZ mode and antisymmetric SPPP hybrid mode. This allows to modulate the polariton resonances over a wide range of frequencies in the mid-IR, from 850 to 4000 cm$^{-1}$.

Besides, we have shown how the SPPP hybrid mode incorporates the advantages of the bare SPP and SPhP modes. The very low damping of the sapphire phonons reduces the overall damping of the hybrid SPPP as compared to the bare SPP, especially for frequencies close to the sapphire reststrahlen band. This has a direct impact on the propagation length of the polariton, a fundamental characteristic for the employment of surface polaritons in nanophotonic devices. For CdZnO thicknesses of 65 nm and 25 nm, significant propagation lengths of 458 µm and 513 µm, respectively, have been obtained, i.e., one order of magnitude higher than that of typical SPPs.

## Acknowledgments

This work was funded by the Spanish Ministry of Economy and Competitiveness (MINECO) through the Project TEC2017-85912-C2, and the Generalitat Valenciana under the project Prometeo II 2015/004. The authors would like to thank Dr. J. Pedrós and R. Fandan for valuable discussions. J.T.-A. holds a Predoctoral Contract from the Universidad Politécnica de Madrid.

**References**

(1) Yang, Y.; Kelley, K.; Sachet, E.; Campione, S.; Luk, T. S.; Maria, J. P.; Sinclair, M. B.; Brener, I. Femtosecond Optical Polarization Switching Using a Cadmium Oxide-Based Perfect Absorber. *Nat. Photonics* **2017**, *11* (6), 390–395.

(2) Luk, T. S.; Campione, S.; Kim, I.; Feng, S.; Jun, Y. C.; Liu, S.; Wright, J. B.; Brener, I.; Catrysse, P. B.; Fan, S.; et al. Directional Perfect Absorption Using Deep Subwavelength Low-Permittivity Films. *Phys. Rev. B - Condens. Matter Mater. Phys.* **2014**, *90* (8), 1–10.

(3) De Hoogh, A.; Opheij, A.; Wulf, M.; Rotenberg, N.; Kuipers, L. Harmonics Generation by Surface Plasmon Polaritons on Single Nanowires. *ACS Photonics* **2016**, *3* (8), 1446–1452.

(4) Fang, Y.; Sun, M. Nanoplasmonic Waveguides: Towards Applications in Integrated Nanophotonic Circuits. *Light Sci. Appl.* **2015**, *4* (December 2014), 1–11.

(5) Gramotnev, D. K.; Bozhevolnyi, S. I. Plasmonics beyond the Diffraction Limit. *Nat. Photonics* **2010**, *4* (2), 83–91.

(6) Vassant, S.; Hugonin, J.-P.; Marquier, F.; Greffet, J.-J. Berreman Mode and Epsilon near Zero Mode. *Opt. Express* **2012**, *20* (21), 23971.

(7) Campione, S.; Brener, I.; Marquier, F. Theory of Epsilon-near-Zero Modes in

Ultrathin Films. *Phys. Rev. B - Condens. Matter Mater. Phys.* **2015**, *91* (12), 1–5.

(8) Boltasseva, A.; Atwater, H. A. Low-Loss Plasmonic Metamaterials. *Science (80-.).* **2011**, *331* (6015), 290–291.

(9) Rhodes, C.; Cerruti, M.; Efremenko, A.; Losego, M.; Aspnes, D. E.; Maria, J.-P.; Franzen, S. Dependence of Plasmon Polaritons on the Thickness of Indium Tin Oxide Thin Films. *J. Appl. Phys.* **2008**, *103* (9), 093108.

(10) Montes Bajo, M.; Tamayo-Arriola, J.; Hugues, M.; Ulloa, J. M.; Le Biavan, N.; Peretti, R.; Julien, F. H.; Faist, J.; Chauveau, J.-M.; Hierro, A. Multisubband Plasmons in Doped ZnO Quantum Wells. *Phys. Rev. Appl.* **2018**, *10* (2), 024005.

(11) Jefferson, P. H.; Hatfield, S. A.; Veal, T. D.; King, P. D. C.; McConville, C. F.; Zúñiga–Pérez, J.; Muñoz–Sanjosé, V. Bandgap and Effective Mass of Epitaxial Cadmium Oxide. *Appl. Phys. Lett.* **2008**, *92* (2), 022101.

(12) Vasheghani Farahani, S. K.; Veal, T. D.; King, P. D. C.; Zúñiga-Pérez, J.; Muñoz-Sanjosé, V.; McConville, C. F. Electron Mobility in CdO Films. *J. Appl. Phys.* **2011**, *109* (7), 073712.

(13) Ziabari, A. A.; Ghodsi, F. E. Optoelectronic Studies of Sol–Gel Derived Nanostructured CdO–ZnO Composite Films. *J. Alloys Compd.* **2011**, *509* (35), 8748–8755.

(14) Sachet, E.; Shelton, C. T.; Harris, J. S.; Gaddy, B. E.; Irving, D. L.; Curtarolo, S.; Donovan, B. F.; Hopkins, P. E.; Sharma, P. A.; Sharma, A. L.; et al. Dysprosium-Doped Cadmium Oxide as a Gateway Material for Mid-Infrared Plasmonics. *Nat. Mater.* **2015**, *14* (4), 414–420.

(15) Wang, A.; Babcock, J. R.; Edleman, N. L.; Metz, A. W.; Lane, M. A.; Asahi, R.; Dravid, V. P.; Kannewurf, C. R.; Freeman, A. J.; Marks, T. J. Indium-Cadmium-Oxide Films Having Exceptional Electrical Conductivity and Optical Transparency: Clues for Optimizing Transparent Conductors. *Proc. Natl. Acad. Sci.* **2001**, *98* (13), 7113–7116.

(16) Yan, M.; Lane, M.; Kannewurf, C. R.; Chang, R. P. H. Highly Conductive Epitaxial Cdo Thin Films Prepared by Pulsed Laser Deposition. *Appl. Phys. Lett.* **2001**, *78* (16), 2342–2344.

(17) Runnerstrom, E. L.; Kelley, K. P.; Sachet, E.; Shelton, C. T.; Maria, J.-P. Epsilon-near-Zero Modes and Surface Plasmon Resonance in Fluorine-Doped Cadmium Oxide Thin Films. *ACS Photonics* **2017**, *4* (8), 1885–1892.

(18) Tamayo-Arriola, J.; Huerta-Barberà, A.; Montes Bajo, M.; Muñoz, E.; Muñoz-Sanjosé, V.; Hierro, A. Rock-Salt CdZnO as a Transparent Conductive Oxide. *Appl. Phys. Lett.* **2018**, *113* (22), 222101.

(19) Nakayama, M. Theory of Surface Waves Coupled to Surface Carriers. *J. Phys. Soc. Japan* **1974**, *36* (2), 393–398.

(20) Brar, V. W.; Jang, M. S.; Sherrott, M.; Kim, S.; Lopez, J. J.; Kim, L. B.; Choi, M.; Atwater, H. Hybrid Surface-Phonon-Plasmon Polariton Modes in Graphene/

Monolayer h - BN Heterostructures. *Nano Lett.* **2014**, *14* (7), 3876–3880.

(21) Burke, J. J.; Stegeman, G. I.; Tamir, T. Surface-Polariton-like Waves Guided by Thin, Lossy Metal Films. *Phys. Rev. B* **1986**, *33* (8), 5186–5201.

(22) Smith, L. H.; Taylor, M. C.; Hooper, I. R.; Barnes, W. L. Field Profiles of Coupled Surface Plasmon-Polaritons. *J. Mod. Opt.* **2008**, *55* (18), 2929–2943.

(23) Otto, A. Excitation of Nonradiative Surface Plasma Waves in Silver by the Method of Frustrated Total Reflection. *Zeitschrift für Phys. A Hadron. Nucl.* **1968**, *216* (4), 398–410.

(24) Oskooi, A. F.; Roundy, D.; Ibanescu, M.; Bermel, P.; Joannopoulos, J. D.; Johnson, S. G. MEEP: A Flexible Free-Software Package for Electromagnetic Simulations by the FDTD Method. *Comput. Phys. Commun.* **2010**, *181*, 687–702.

(25) Runnerstrom, E. L.; Kelley, K. P.; Folland, T. G.; Nolen, J. R.; Engheta, N.; Caldwell, J. D.; Maria, J. Polaritonic Hybrid-Epsilon-near-Zero Modes: Beating the Plasmonic Confinement vs Propagation-Length Trade-Off with Doped Cadmium Oxide Bilayers. *Nano Lett.* **2019**, *19* (2), 948–957.

(26) Passler, N. C.; Gubbin, C. R.; Folland, T. G.; Razdolski, I.; Katzer, D. S.; Storm, D. F.; Wolf, M.; De Liberato, S.; Caldwell, J. D.; Paarmann, A. Strong Coupling of Epsilon-Near-Zero Phonon Polaritons in Polar Dielectric Heterostructures. *Nano Lett.* **2018**, *18* (7), 4285–4292.

(27) Barnes, W. L.; Dereux, A.; Ebbesen, T. W. Surface Plasmon Subwavelength Optics. *Nature* **2003**, *424* (6950), 824–830.

# Supporting Information

Section 1 shows the details of the infrared reflectance measurements and modelling carried out to derive the fundamental optical parameters of the CdZnO films and sapphire substrate. In Section 2 the ATR geometry and the experimentally obtained and modelled ATR curves for all samples are shown. In Section 3, the procedure to obtain the real and imaginary parts of the in-plane and out-of-plane momentum of the hybrid mode is described.

1. **Infrared reflectance spectroscopy: experimental and modelling**

The infrared reflectance spectra of the samples were measured and fitted in order to obtain the fundamental parameters of the CdZnO films and the sapphire substrate: the CdZnO plasma frequency ($\omega_p$) and its damping ($\gamma_p$), the CdZnO film thickness ($a$) and the sapphire phonon modes. To do so, the dielectric function of each constituent layer was modelled and the Transfer Matrix Method (TMM) was applied.

First, a *r*-plane sapphire substrate was measured and fitted and its phonon modes were derived. Both in the reflectance and attenuated total reflectance (ATR) measurements, the sapphire substrate was oriented in the direction where the axial phonons are minimized,[1] and therefore to fit the reflectance and ATR spectra only the planar phonons are required.

The contribution of the sapphire phonons to the dielectric function are considered through Gervais harmonic oscillators, as

$$\varepsilon(\omega) = \varepsilon_\infty^{Sapphire} \prod_j \frac{\omega_{LO,j}^2 - \omega^2 - i\gamma_{LO,j}\omega}{\omega_{TO,j}^2 - \omega^2 - i\gamma_{TO,j}\omega}$$

where the product runs through all the planar phonon modes found in sapphire, $\varepsilon_\infty^{Sapphire}$ is the high-frequency dielectric constant of sapphire, $\omega_{TO,LO}$ are the frequencies of the transversal optical (TO) and longitudinal optical (LO) phonons, respectively, and $\gamma_{TO,LO}$ the damping of the TO and LO oscillators.

The measured and modelled reflectance curves of the *r*-plane sapphire substrate in *p*-polarization at an incidence angle of the infrared light of 45° are shown in Figure S1. The extracted values of the phonon frequencies are listed in Table S1.

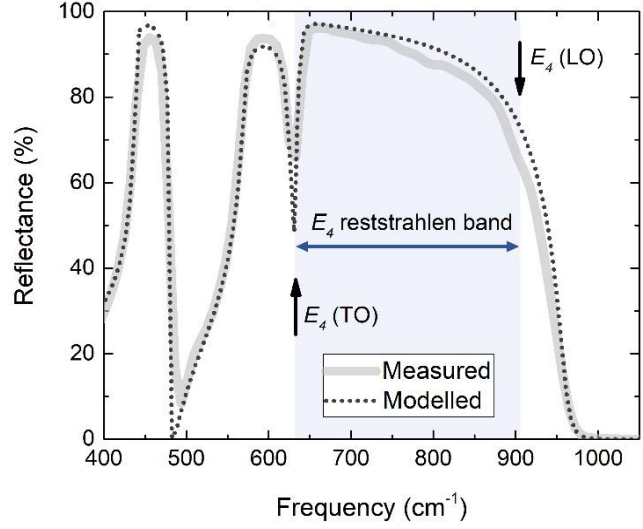

**Figure S1.** Reflectance spectrum (continuous line) and fitted model (dotted line) of the bare *r*-plane sapphire substrate at 45° in *p*-polarization. The shaded region indicates the reststrahlen band of the $E_4$ mode, where the sapphire surface phonon is formed.

| Phonon mode | $\omega_{LO}$ (cm⁻¹) | $\gamma_{LO}$ (cm⁻¹) | $\omega_{TO}$ (cm⁻¹) | $\gamma_{TO}$ (cm⁻¹) |
|---|---|---|---|---|
| $E_1$ | 388 | 4 | 385 | 4 |
| $E_2$ | 481 | 1 | 440 | 1 |
| $E_3$ | 629 | 7 | 570 | 5 |
| $E_4$ | 906 | 17 | 633 | 6 |

**Table S1.** Frequencies and dampings of the phonon modes of the *r*-plane sapphire substrate derived from infrared reflectance spectroscopy. $E_4$ is the phonon mode interacting with the CdZnO surface plasmons.

Apart from the values shown in Table S1, $\varepsilon_\infty^{Sapphire}$ was deduced to be 3.0. These results are in excellent agreement with those by Schubert *et al.*[1] and are used as input parameters for modelling the reflectance spectra of the CdZnO-sapphire films.

On the other hand, the dielectric function of CdZnO has to account with the interaction of the infrared light with the free electrons through the Drude model, as

$$\varepsilon(\omega) = \varepsilon_\infty^{CdZnO}\left(1 - \frac{\omega_p^2}{\omega^2 - i\gamma_p\omega}\right).$$

Here, the plasma frequency is defined as

$$\omega_p = \sqrt{\frac{ne^2}{m_e^* \varepsilon_0 \varepsilon_\infty^{CdZnO}}},$$

where $n$ is the free electron concentration, $e$ the electron charge, $m_e^*$ the electron effective mass, $\varepsilon_0$ the vacuum permittivity, and $\varepsilon_\infty^{CdZnO}$ is the high-frequency dielectric constant of CdZnO.

The measured and modelled reflectance curves for all the samples are shown in Figure S2, and the extracted values from the fits are shown in Table S2.

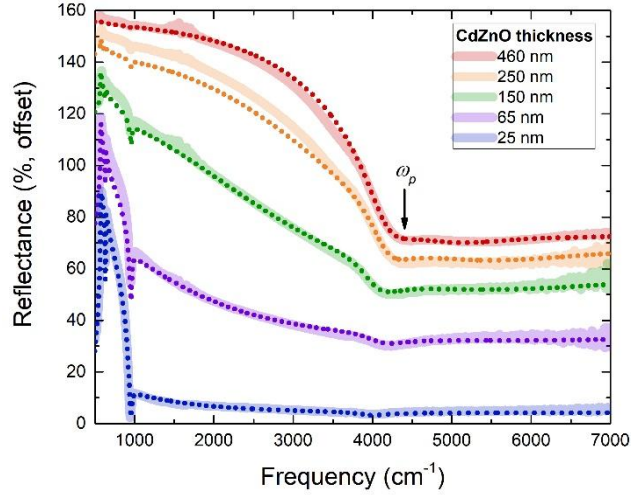

**Figure S2.** Reflectance spectra (continuous lines) and the fitted model (dotted lines) of all the samples at 45° in *p*-polarization, offset for clarity. The plasma frequency of the thickest CdZnO film is indicated as an example, falling at the dip in the reflectance spectrum.

| $a$ (nm) | $\omega_p$ (cm$^{-1}$) | $\gamma_p$ (cm$^{-1}$) |
|---|---|---|
| 460 | 4082 | 525 |
| 250 | 4121 | 518 |
| 150 | 4043 | 518 |
| 65  | 4098 | 550 |
| 25  | 3980 | 503 |

**Table S2.** Values of the fitted parameters of the CdZnO films from reflectance spectroscopy. A value of $\varepsilon_\infty^{CdZnO} = 5.1$ was obtained for all films.

As can be observed in Table S2, except of the thickness, all the other fitted parameters are very similar in all the CdZnO films. Therefore, essentially all the changes in the reflectance spectra arise from the variation of the CdZnO thickness. Besides, it is worth to note the difference between the damping of the CdZnO plasmons and that of sapphire phonons. While in CdZnO the plasmons have dampings around 520 cm$^{-1}$, the $E_4$ phonon mode of sapphire has a damping of 6 cm$^{-1}$ for the TO and 17 cm$^{-1}$ for the LO. Thus, the damping of the SPPP hybrid mode must be between these two limits, depending on the proximity of the resonance to the sapphire phonons.

Finally, all the parameters derived from these fits were used as input parameters for the ATR simulations, in order to compare the modelled ATR curves with the experimental ones.

## 2. ATR measurements and simulations

In order to excite the surface modes, ATR measurements were carried out in the Otto configuration,[2] as schematized in Figure S3.

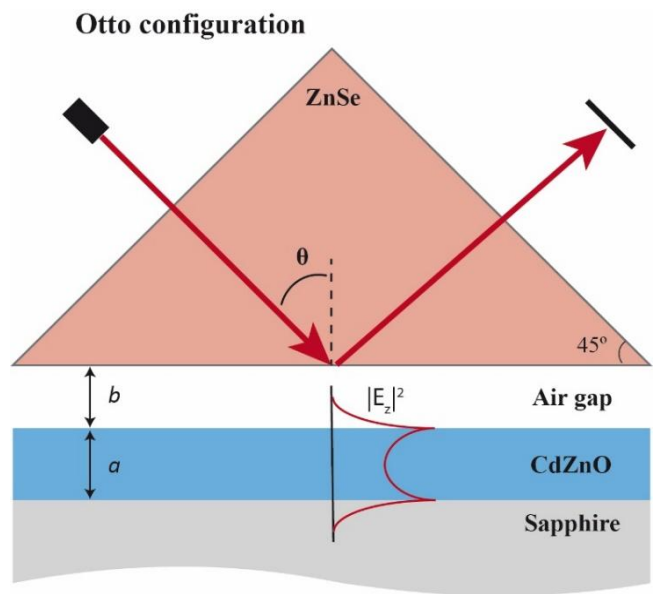

**Figure S3.** Schematic of the ATR measurements carried out in the Otto configuration. The evanescent wave generated at the prism-air interface excites the surface modes of the air-CdZnO-sapphire system. The $z$ component of the electric field of the symmetric mode is represented.

A ZnSe prism with an internal angle of 45° and a refractive index of 2.37 is used, and the ATR curves were taken in $p$-polarization, varying the incidence angle ($\theta$) in the range between 39° and 59°. Thus, the in-plane momentum is matched to the polariton momentum by changing the incidence angle, and the resonance frequency of the surface plasmon polariton (SPP) and the surface plasmon phonon polariton (SPPP) shift accordingly.

The measured ATR curves of all samples are shown in Figure S4, and, in order to compare them with those modelled with the TMM, the simulated ATR curves are shown in Figure S5. The parameters for modelling the ATR curves are taken from the reflectance measurements described in Section 1, and the only parameter to be fitted is the air gap, which results to be about 350 nm in all the experiments. In each case, the upper branch (UB), corresponding to the symmetric mode, and the lower branch (LB) corresponding to the antisymmetric mode, are indicated.

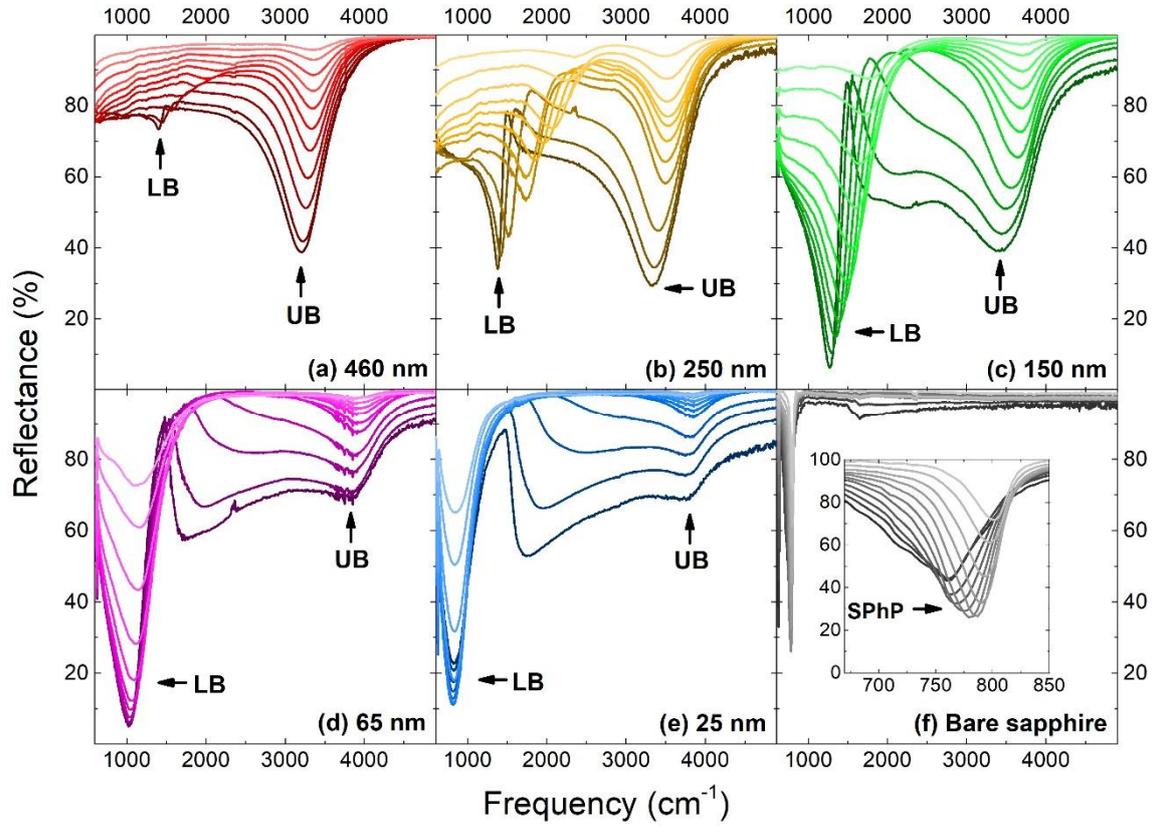

**Figure S4.** From (a) to (d), ATR measured curves for decreasing CdZnO thickness. The incidence angle was varied by 2° steps in the range from 39° (dark colors) to 59° (light colors). (f) ATR measured curve of the bare *r*-plane sapphire substrate. The inset shows in more detail the sapphire surface phonon polariton.

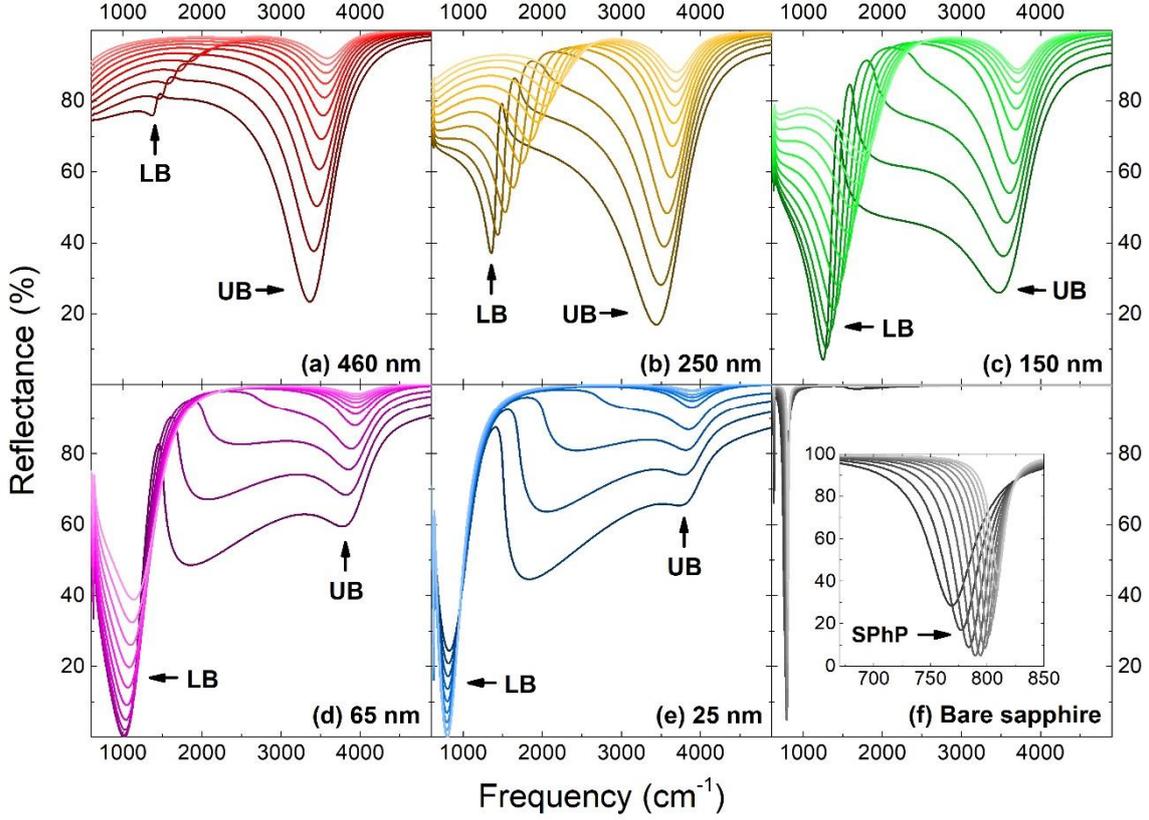

**Figure S5.** From (a) to (d), ATR simulated curves with the TMM for decreasing CdZnO thickness. The incidence angle was varied by 2° steps in the range from 39° (dark colors) to 59° (light colors). (f) ATR simulated curve of the bare *r*-plane sapphire substrate. The inset shows in more detail the sapphire surface phonon polariton.

## 3. Derivation of the real and imaginary parts of the in-plane and out-of plane momenta of the hybrid SPPP

Assessing the real and imaginary parts of the in-plane and out-of-plane momenta of the surface modes is key to compare the experimentally obtained resonance frequencies of the SPPP with those predicted by the theory, and to compute its propagation length ($L_p$) and confinement distance ($\delta_z$). While the air-CdZnO-sapphire system is enough to qualitatively describe the nature of the surface polaritons, in order to accurately derive the dispersion curves and $L_p$ and $\delta_z$ of the SPPP the prism has also to be considered, i.e. air is considered with a finite thickness, and the ZnSe prism is treated as a semi-infinite medium.

Following the derivation deduced by Baltar *et al.*,[3] $Re\{k_x, k_z\}$ and $Im\{k_x, k_z\}$ in a four-layer system can be obtained from

$$\det \begin{vmatrix} R_1 e^{-\gamma_2(a+b)} - R_2 e^{-\gamma_2(a+b)} & R_1 e^{\gamma_2(a+b)} + R_2 e^{\gamma_2(a+b)} & 0 & 0 \\ 0 & 0 & R_4 + R_3 & R_4 - R_3 \\ e^{-\gamma_2 a} & e^{\gamma_2 a} & -e^{-\gamma_3 a} & -e^{\gamma_3 a} \\ R_2 e^{-\gamma_2 a} & -R_2 e^{\gamma_2 a} & -R_3 e^{-\gamma_3 a} & R_3 e^{\gamma_3 a} \end{vmatrix} = 0.$$

Here, the zeros of the determinant define the implicit relation between $Re\{k_x, k_z\}$ and $Im\{k_x, k_z\}$ with the frequency of the incident light. Letters *a* and *b* indicate the thickness of the CdZnO and the air gap, respectively, and the subscript numbers are related to the involved materials: 1 for the ZnSe prism, 2 for air, 3 for CdZnO and 4 for sapphire. The different parameters are defined as:[3]

$$\gamma = ik_z = \sqrt{k_x^2 - \varepsilon k_0^2} \quad \text{with} \quad k_0 = \omega/c,$$

where *c* is the speed of light, and

$$R_i = \gamma_i/\varepsilon_i,$$

where the subscript *i* takes the number of the corresponding layer and $\varepsilon_i$ is its associated dielectric function.

**References**

[1] M. Schubert, T.E. Tiwald, and C.M. Herzinger, Phys. Rev. B **61**, 8187 (2000).

[2] A. Otto, Zeitschrift Für Phys. A Hadron. Nucl. **216**, 398 (1968).

[3] H.T.M.C.M. Baltar, K. Drozdowicz-tomsia, and E.M. Goldys, *Plasmonics - Principles and Applications* (InTech, 2012).